\documentclass[twocolumn,showpacs,preprintnumbers,amsmath,amssymb,floatfix]{revtex4}

\usepackage{graphicx}
\usepackage{dcolumn}
\usepackage{bm}

\newcommand{\pom}{\tt I\! P}
\newcommand{\beq}{\begin{equation}}
\newcommand{\eeq}{\end{equation}}

\begin{document}

\title{An estimation of single and double diffractive heavy flavour production in hadron-hadron colliders}

\author{M. V. T. Machado}

\affiliation{Centro de Ci\^encias Exatas e Tecnol\'ogicas, Universidade Federal do Pampa \\
Campus de Bag\'e, Rua Carlos Barbosa. CEP 96400-970. Bag\'e, RS, Brazil}

\begin{abstract}

Results from a phenomenological analysis for diffractive
hadroproduction of heavy flavors at high energies are reported. Diffractive production of charm, bottom and top are calculated using Regge factorization, taking into account recent experimental determination of the diffractive parton density functions in Pomeron by the H1 Collaboration at DESY-HERA. In addition, multiple-Pomeron corrections are considered through the rapidity gap survival probability factor. We give numerical predictions for single diffractive as well as double Pomeron exchange (DPE) cross sections, which agree with the available data for diffractive production of charm and beauty.  We make estimates which could be compared to future measurements at the LHC.

\end{abstract}


\maketitle

\section{Introduction}

Diffractive processes in hadron collisions are well described, with respect to the overall cross-sections, by Regge theory in terms of the exchange of a Pomeron  with vacuum quantum numbers \cite{Collins}. However, the nature of the Pomeron and its reaction mechanisms are not completely known. A good channel attracting much attention is the use of hard scattering to resolve the quark and gluon content in the Pomeron \cite{IS}. Such a parton structure is natural in a modern QCD approach to the strongly interacting Pomeron. The systematic observations of diffractive deep inelastic scattering (DDIS) at HERA have increased the knowledge about the QCD Pomeron, providing us with the diffractive distributions of singlet quarks and gluons in Pomeron as well as the diffractive structure function \cite{H1diff}. In hadronic collisions, we shall characterize an event as single diffractive if one of the colliding hadrons emits a Pomeron that scatters off the other hadron. A single diffractive reaction would therefore correspond to, $h_i+h_j\rightarrow h_i+X$, in which the intact hadron $i$ emits a Pomeron ($h_i\rightarrow h_i+\pom$) which interacts with hadron $j$ ($h_j + \pom \rightarrow X$) leading to the referred reaction. Hadron $i$ is detected in the final state with a large longitudinal momentum fraction. Hard diffractive events with a large momentum transfer are also characterized by the absence of hadronic energy in certain angular regions of the final state phase space (rapidity gaps). The events fulfilling the conditions of large rapidity gaps and a highly excited hadron remnant are named single diffractive in contrast to those in which both colliding hadrons remain intact as they each emit a Pomeron (central diffraction or double Pomeron exchange events). At high energies, there are important
contributions from unitarization effects to the single-Pomeron exchange cross section. These absorptive or unitarity corrections cause the suppression of any large rapidity gap process, except elastic scattering. In the black disk limit the absorptive corrections may completely terminate those processes. This partially occurs in (anti)proton--proton collisions, where unitarity is nearly saturated at small impact parameters \cite{k3p}.  The multiple-Pomeron
contributions depend, in general, on the particular
hard process and it is called survival probability factor.  At the Tevatron energy, $\sqrt s = 1.8$~TeV, the
suppression is of order 0.05--0.2~\cite{GLM,KMRsoft,BH,KKMR}, whereas for LHC energy, $\sqrt{s}=14$ TeV, the suppression  appears to be 0.08--0.1 ~\cite{GLM,KMRsoft,KKMR}. These corrections are, therefore, crucial for the reliability of the theoretical predictions for hard diffractive processes.

We present below a calculation for diffractive production of heavy quarks in proton-proton collisions. The motivation is to produce updated theoretical estimations compatible with the scarce accelerator data on single diffractive charm and bottom hadroproduction \cite{sdccdata,sdbbdata} and to obtain reliable predictions to the future measurements at the LHC. These predictions are quite important in the determination of the background processes for diffractive Higgs production and related reactions. The background for Higgs is a subject of intense debate in literature and is out of the scope of present work. Let us illustrate two representative examples: (a) for a low mass Higgs, $M_H\leq 150$ GeV, an interesting exclusive channel is the $b\bar{b}$ decay mode $pp \rightarrow p \,(H\rightarrow b\bar{b})\,p$ \cite{hbbmode}; (b) the inclusive  channel $pp\rightarrow (H\rightarrow WW \rightarrow l^+ l^- /\!\!\!p_T )\,X$ \cite{wwmode}. In both cases the heavy quark background contribution is sizable \cite{hbbmode,wwmode}, namely the DPE bottom production in (a) and lepton contribution from heavy quark decays in (b).

For the present purpose we rely on the Regge factorization and the corresponding corrections for multiple-Pomeron scatterings. Factorization for diffractive hard scattering is equivalent to the hard-scattering aspects of the Ingelman and Schlein model \cite{IS}, where diffractive scattering is attributed to the
exchange of a Pomeron, i.e. a colorless object with vacuum quantum
numbers. The Pomeron is treated like a real particle. Thus, one considers that a diffractive electron-proton collision is
due to an electron-Pomeron collision. Similarly, a diffractive
proton-proton collision occurs due to a proton-Pomeron collision.
Therefore, the diffractive hard cross sections are
obtained as a product of a hard-scattering coefficient, a known Pomeron-proton coupling, and parton densities in
the Pomeron. The parton densities in the Pomeron have been systematically extracted from
diffractive DIS measurements. In particular, the quark singlet and gluon content of the Pomeron is obtained from the diffractive structure function $F_{2}^{D(3)}(x_{\pom},\beta,Q^2)$.  Recently, a new analysis of these diffractive parton distributions has been presented \cite{H1diff} by the H1 Collaboration in DESY-HERA.

The paper is organized as follows. In the next section, we present the main formulae to compute the inclusive and diffractive  cross sections (single and central diffraction) for heavy flavors hadroproduction. We show the details concerning the parameterization for the diffractive partons distribution in the Pomeron. In addition, we present the theoretical estimations for the rapidity gap survival probability factor.  In the last section we present the numerical results, taking properly the experimental cuts, and perform predictions to future measurements in the CERN LHC experiment. The compatibility with data is analyzed and the comparison with other approaches is considered.

%
%
\section{Diffractive Hadroproduction of Heavy Flavors}

Let us start by introducing the main expressions to compute the inclusive and diffractive cross sections for heavy flavor production in hadron colliders. The starting point is the inclusive total cross section for a process in which partons of two hadrons, $h_1$ and $h_2$, interact to produce a heavy quark pair, $h_1 + h_2 \rightarrow Q\bar{Q} + X$, at center of mass energy $\sqrt{s}$. At leading order (LO) heavy quarks are produced by $gg$ fusion and
$q \overline q$ annihilation while at next-to-leading order (NLO), $qg +
\overline q g$ scattering is also included.  At any order, the partonic
cross section may be expressed in terms of dimensionless scaling functions
$f^{(k,l)}_{ij}$ that depend only on the variable $\rho$ \cite{MNR},
\begin{eqnarray}
\label{scalingfunctions}
\hat \sigma_{ij}(\hat s,m_Q^2,\mu_F^2,\mu_R^2) & = & \frac{\alpha^2_s(\mu_R)}{m_Q^2}
\sum\limits_{k=0}^{\infty} \,\, \left[ 4 \pi \alpha_s(\mu_R) \right]^k\nonumber\\
& \times & \sum\limits_{l=0}^k \,\, f^{(k,l)}_{ij}(\rho) \,\,
\ln^l\left(\frac{\mu_F^2}{m_Q^2}\right) \, ,
\end{eqnarray}
where $\hat s$ is the partonic center of mass, $m_Q$ is the heavy quark mass,
$\mu_R$ ($\mu_F$) is the renormalization (factorization) scale, and $\rho = \hat s/4 m_Q^2 - 1$.
The cross section is calculated as an expansion in powers of $\alpha_s$
with $k=0$ corresponding to the Born cross section at order ${\cal
O}(\alpha_s^2)$.  The first correction, $k=1$, corresponds to the NLO cross
section at ${\cal O}(\alpha_s^3)$.  The total hadronic cross section is obtained by convoluting the total partonic
cross section with the parton distribution functions of the initial hadrons,
\begin{eqnarray}
\label{totalhadroncrs}
\sigma_{h_1h_2}(s,m_Q^2) \!\! & = & \!\! \sum_{i,j} 
\int_{\tau}^1dx_1\int_{\frac{\tau}{x_1}}^1 dx_2\,f_i^{h_1}(x_1,\mu_F^2) f_j^{h_2}(x_2,\mu_F^2) \nonumber \\
&\times & \hat \sigma_{ij}(\hat{s} ,m_Q^2,\mu_F^2,\mu_R^2)\, ,
\end{eqnarray}
where the sum $i,j = q,\bar{q},g$ is over all massless partons,
$x_1$ and $x_2$ are the  hadron momentum fractions carried by
the interacting partons and  $\tau=4m_Q^2/s$. The parton distribution functions, denoted by $f_i^p(x_i,\mu)$, are evaluated at
the factorization scale, assumed to be equal to the renormalization scale
in  our calculations.

\begin{figure}[t]
\includegraphics[scale=0.45]{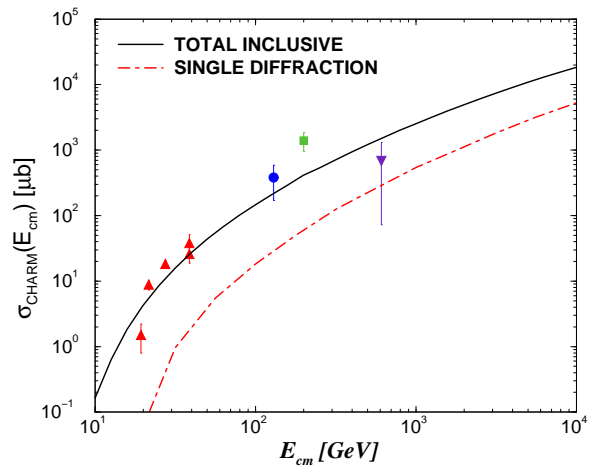}
\caption{(Color online) Total cross section for inclusive charm hadroproduction (solid line) and  corresponding  single diffractive cross section without multiple-Pomeron correction (dot-dashed line). Available accelerator data are also shown (see text).}
\label{fig:1}
\end{figure}

\begin{figure}[t]
\includegraphics[scale=0.45]{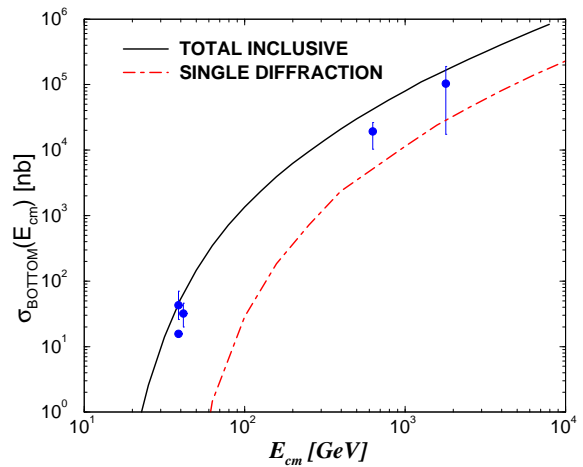}
\caption{(Color online) Total cross section for inclusive bottom hadroproduction (solid line) and  corresponding  single diffractive cross section without multiple-Pomeron correction (dot-dashed line). Available accelerator data are also shown (see text).}
\label{fig:2}
\end{figure}

For sake of illustration, we calculate numerically the total inclusive cross sections for heavy flavors pair-production using the MRST set of partons \cite{mrst2004nlo}. The low $x$ region is particularly relevant for $Q \overline Q$ production at the LHC as well as at Tevatron. For charm and bottom production, the $gg$ process becomes dominant and information on the gluon distribution is of particular importance. We compute the cross sections with the following mass and
scale parameters: $\mu_c = 2m_c$ (with $m_c = 1.5$ GeV) and $\mu_b = m_b$ (with $m_b = 4.5$ GeV).  For the top quark case, we use $\mu_t = m_t$ (where $m_t=176$ GeV). The present choice for the scales is based on the current phenomenology for heavy quark hadroproduction \cite{Ramonahq}. The agreement with the total charm cross section data is fairly good. In Figs. \ref{fig:1} and \ref{fig:2} the numerical results (solid curves) are compared to accelerator data of charm and bottom hadroproduction \cite{hqdata}, respectively. For the top cross section, we obtain $\sigma_{tot}(\sqrt{s}=1.8\,\mathrm{TeV})=5$ pb and $\sigma_{tot}(\sqrt{s}=14\,\mathrm{TeV})=2$ nb. It should be stressed that sizable uncertainties are introduced by changing, for instance, quark masses and/or the renormalization scale. However, our purpose here is to estimate the diffractive ratios $\sigma^D/\sigma_{tot}$, which are less sensitive to a particular choice.

For the {\it hard} diffractive
processes we will consider the Ingelman-Schlein (IS) picture \cite{IS}, where
the Pomeron structure (quark and gluon content) is probed. In the case of single diffraction, a Pomeron is emitted by one of the colliding hadrons. That hadron is detected, at least in principle, in the final state and the remaining hadron scatters off the emitted Pomeron. A typical single diffractive reaction is given by $p+p\rightarrow p+Q\bar{Q}+X$. In the IS approach, the  single diffractive cross section  is assumed to factorise into the total Pomeron--hadron cross
section and the  Pomeron  flux  factor  \cite{IS}.  The single
diffractive event may then be written as (for equal hadrons)
\begin{eqnarray}
 \label{sdexp}
\frac{d\sigma^{\mathrm{SD}}\,(hh\rightarrow h+  Q\bar{Q}+X)}
{dx^{(i)}_{\pom}d|t_i|}\! & = & \! f_{{\rm\pom}/i}(x^{(i)}_{\pom},|t_i|) \nonumber\\
\! & \times & \!\sigma\left({\pom} + h\rightarrow  Q\bar{Q}  +  X\right),\nonumber
\end{eqnarray}
where the Pomeron kinematical variable $x_{\pom}$ is defined as $x_{\pom}^{(i)}=s_{\pom}^{(j)}/s_{ij}$, where  $\sqrt{s_{\pom}^{(j)}}$ is the center-of-mass energy in the Pomeron--hadron $j$ system and $\sqrt{s_{ij}}=\sqrt{s}$ the center-of-mass energy in the hadron $i$--hadron $j$ system. The momentum transfer in the  hadron $i$ vertex is denoted by $t_i$. A similar factorization can also be applied to double Pomeron exchange (DPE) process, where both colliding
hadrons  can in  principle  be detected in the final  state. This diffractive process is also known as central diffraction (CD). Thus, a
typical reaction would be  $p+p \rightarrow p+p
+ Q\bar{Q} + X$, and DPE  events thus are characterized by
two  quasi--elastic hadrons with  rapidity  gaps between them and the
central heavy flavor products. The DPE cross section may then be written as,
\begin{eqnarray}
 \label{sddexp}
\frac{d\sigma^{\mathrm{DD}}\,(hh\rightarrow h+ h+ Q\bar{Q}+X)}
{dx^{(i)}_{\pom}dx^{(j)}_{\pom}d|t_i|d|t_j|}\!\! & = & \!\! f_{{\rm\pom}/i}(x^{(i)}_{\pom},|t_i|)\, f_{{\rm\pom}/j}(x^{(j)}_{\pom},|t_j|) \nonumber\\
\! & \times & \!\sigma\left({\pom} + {\pom}\rightarrow  Q\bar{Q}  +  X\right),\nonumber
\end{eqnarray}

In order to obtain the corresponding expression for
diffractive processes, one assumes that one of the hadrons, say
hadron $h_1$, emits a Pomeron whose partons interact with partons of the hadron $h_2$.
Thus the parton distribution  $x_1 f_{i/h_1}(x_1, \mu^2)$ in
Eq.~(\ref{totalhadroncrs}) is replaced by the convolution between a putative
distribution of partons in the Pomeron, $\beta f_{a/{\pom}}(\beta,
\mu^2)$, and the ``emission rate" of Pomerons by the hadron, $f_{{\pom}/h}(x_{{\pom}},t)$. The last quantity, $f_{{\pom}/h}(x_{{\pom}},t)$, is the Pomeron flux factor and its explicit formulation is described in
terms of Regge theory. Therefore, we can rewrite the parton distribution as
\begin{eqnarray}
\nonumber
\label{convol}
x_1 f_{a/h_1}(x_1, \,\mu^2) & =& \int dx_{{\pom}} \int d\beta \int dt\,
f_{{\pom}/h_1}(x_{{\pom}},\,t) \\
&\times & \beta \, f_{a/{\pom}}(\beta, \,\mu^2)\,
\delta \left(\beta-\frac{x_1}{x_{{\pom}}}\right),
\end{eqnarray}
and, now defining $\bar{f} (x_{{\pom}}) \equiv \int_{-\infty}^0 dt\
f_{{\pom/h_1}}(x_{{\pom}},t)$, one obtains
\begin{eqnarray}
\label{convoP}
x_1 f_{a/h_1}(x_1, \,\mu^2)\ =\ \int dx_{{\pom}} \
\bar{f}(x_{{\pom}})\, {\frac{x_1}{x_{{\pom}}}}\, f_{a/{\pom}}
({\frac{x_1}{x_{{\pom}}}}, \mu^2).
\end{eqnarray}

Using  the  substitution  given  in  Eq.~(\ref{convoP})  we  can
proceed in write  down  the   expressions   for  the  single  and  central (DPE) diffractive     cross    sections    for     $Q\bar{Q}$     production using the expression for the total cross section in Eq. ({\ref{totalhadroncrs}),
\begin{eqnarray}
\label{sdxsect}
& &\!\sigma_{h_1h_2}^{\mathrm{SD}}(s,m_Q^2)   =   \sum_{i,j=q\bar{q},g} 
\int_{\tau}^1dx_1\int_{\tau/x_1}^1 dx_2\int_{x_1}^{x_{\pom}^{\mathrm{max}}}\frac{dx_{\pom}^{(1)}}{x_{\pom}^{(1)}}\nonumber \\
& \times & \!\bar{f}_{\pom/h_1}\left(x_{\pom}^{(1)}\right)f_{i/\pom}\left(\frac{x_1}{x_{\pom}^{(1)}},\mu^2\right)  f_{j/h_2}(x_2,\mu^2)\,
\hat \sigma_{ij}(\hat{s} ,m_Q^2,\mu^2) \nonumber\\
& & +\, (1\rightleftharpoons 2)\,.
\end{eqnarray}

Similar expression holds for the DPE process, which reads as,
\begin{eqnarray}
\label{ddxsect}
& &\!\sigma_{h_1h_2}^{\mathrm{DPE}}(s,m_Q^2)   =   \sum_{i,j=q\bar{q},g} 
\int_{\tau}^1dx_1\int_{\tau/x_1}^1 dx_2\nonumber \\
& \times & \int_{x_1}^{x_{\pom}^{\mathrm{max}}}\frac{dx_{\pom}^{(1)}}{x_{\pom}^{(1)}} \int_{x_2}^{x_{\pom}^{\mathrm{max}}}\frac{dx_{\pom}^{(2)}}{x_{\pom}^{(2)}} \bar{f}_{\pom/h_1}\left(x_{\pom}^{(1)}\right)\bar{f}_{\pom/h_2}\left(x_{\pom}^{(2)}\right) \nonumber\\
& \times & f_{i/\pom}\left(\frac{x_1}{x_{\pom}^{(1)}},\mu^2\right)\,  f_{j/\pom}\left(\frac{x_2}{x_{\pom}^{(2)}},\mu^2\right)\, 
\hat \sigma_{ij}(\hat{s} ,m_Q^2,\mu^2).
\end{eqnarray}

Having presented the main expressions for single and central diffraction, in the following subsections we shortly present some details on the Pomeron structure function and Pomeron flux that will be employed in our phenomenological studies. In addition, an estimation for the multiple-Pomeron corrections is discussed.

\subsection{The Pomeron Structure Function}

In the estimates for the diffractive cross sections, we will consider the diffractive pdf's recently obtained by the H1 Collaboration at DESY-HERA \cite{H1diff}. The Pomeron structure function has been modeled in terms of a
light flavour singlet distribution $\Sigma(z)$, consisting of $u$, $d$ and $s$
quarks and anti-quarks
with $u=d=s=\bar{u}=\bar{d}=\bar{s}$,
and a gluon distribution $g(z)$.  Here, $z$ is the longitudinal momentum
fraction of the parton entering the hard sub-process
with respect to the diffractive
exchange, such that $z=\beta$
for the lowest order quark-parton model process,
whereas $0<\beta<z$ for higher order processes. The Pomeron carries vacuum quantum numbers, thus it is assumed that the Pomeron quark and antiquark distributions are equal and flavour independent: $q_{\pom}^f=\bar{q}_{\pom}^f=\frac{1}{2N_f}\Sigma_{\pom}$, where $\Sigma_{\pom}$ is a Pomeron singlet quark distribution and $N_f$ is the number of active flavours. The quark singlet and gluon distributions are parameterized
at $Q_0^2$ with the general form,
\begin{equation}
z f_i (z,\,Q_0^2) = A_i \, z^{B_i} \, (1 - z)^{C_i} \,\exp\left[{- \frac{0.01}{(1-z)}}\right]\,
\label{param:general}
\end{equation}
where the last exponential factor ensures that the diffractive pdf's vanish
at $z = 1$.  For the quark singlet distribution,
the data require the inclusion of all three parameters
$A_q$, $B_q$ and $C_q$ in equation~\ref{param:general}.
By comparison, the gluon density is weakly
constrained by the data, which is found to be
insensitive to the $B_g$ parameter. The gluon density is thus
parameterized at $Q_0^2$ using only the $A_g$ and $C_g$ parameters.
With this parameterization, one has the value  $Q_0^2 = 1.75 \ {\rm GeV^2}$ and it is referred to as the `H1 2006 DPDF Fit A'. It is verified that the fit procedure is not sensitive to the gluon pdf and a new adjust was done with $C_g=0$. Thus, the gluon density is then a simple constant at the starting scale for evolution, which was chosen to be
$Q_0^2 = 2.5 \ {\rm GeV^2}$ and it is referred to as the
`H1 2006 DPDF Fit B'. 

Another important element in the calculation is the Pomeron flux factor, introduced in
Eq.~(\ref{convol}). We take the experimental analysis of the diffractive structure function \cite{H1diff}, where the $x_{\pom}$ dependence is parameterized using a flux factor
motivated by Regge theory \cite{Collins},
\begin{eqnarray}
f_{\pom/p}(x_{\pom}, t) = A_{\pom} \cdot
\frac{e^{B_{\pom} t}}{x_{\pom}^{2\alpha_{\pom} (t)-1}} \ ,
\label{eq:fluxfac}
\end{eqnarray}
where  the Pomeron trajectory is assumed to be linear,
$\alpha_{\pom} (t)= \alpha_{\pom} (0) + \alpha_{\pom}^\prime t$, and the parameters
$B_{\pom}$ and $\alpha_{\pom}^\prime$ and their uncertainties are obtained from
fits to H1 FPS data \cite{H1FPS}. The normalization parameter $A_{\pom}$ is chosen such that
$x_{\pom} \cdot \int_{t_{\rm cut}}^{t_{\rm min}} f_{\pom/p} (x_{\pom},t) {\rm d} t
= 1$ at $x_{\pom}= 3\cdot 10^{-3}$, where
$|t_{\rm min}| \simeq m_p^2 \, x_{\pom}^2 \, / \, (1 - x_{\pom})$ is the minimum
kinematically accessible value of $|t|$, $m_p$ is the proton mass and
$|t_{\rm cut}|= 1.0 \rm\ GeV^{2}$ is the limit of the measurement. Expression in Eq. (\ref{eq:fluxfac}) corresponds to the standard Pomeron flux from Regge phenomenology, based on the Donnachie-Landshoff model \cite{DLflux}.

\subsection{Corrections for  multiple-Pomeron scattering}

In the following analysis we will consider the suppression of the hard diffractive cross section by multiple-Pomeron scattering effects. This is taken into account through a gap survival probability factor. There has been large interest in the probability of rapidity gaps in high energy interactions to survive as they may be populated by secondary particles generated by rescattering processes. This effect can be described in terms of screening or absorptive corrections \cite{Bj}. This suppression factor of a hard process accompanied by a rapidity gap depends not only on the probability of the initial state survive, but is sensitive to the spatial distribution of partons inside the incoming hadrons, and thus on the dynamics of the whole diffractive part of the scattering matrix.

Let us introduce a short discussion on the usual procedure to determine the survival probability factor, estimation of its size and the uncertainties in its determination. The survival factor of a large rapidity gap (LRG) in a hadronic final state is the probability of a given LRG not be filled by debris, which originate from the soft re-scattering of the spectator partons and/or from the gluon radiation emitted by partons taking part in the hard interaction. Let ${\cal A}(s,b)$ be the amplitude of the particular diffractive process of interest, considered in the impact parameter, $b$, space. Therefore, the probability that there is no extra inelastic interaction is
\begin{eqnarray}
<\!|S|^2\!>= \frac{\int d^2b\,|{\cal A}(s,b)|^2\,\exp \left[-\Omega (s,b)\right]}{\int d^2b\,|{\cal A}(s,b)|^2}\,,
\label{sdef}
\end{eqnarray}
where $\Omega$ is the opacity (or optical density) of the interaction. This quantity can be computed using a simple one-channel eikonal model or more involved multiple-channel model. The opacity $\Omega (s,b)$ reaches a maximum in the centre of proton and becomes small in the periphery. Therefore, the survival factor depends on the spatial distribution of the constituents of the relevant subprocess. For instance, the spatial $b$-distribution of single and double rapidity gap processes are assumed to be controlled by the slope $B$ of the 
Pomeron-proton vertex, $\beta (t)\propto \exp(-B|t|)$, and that there is no shrinkage coming from the Pomeron amplitude associated with the LRG in hard diffractive subprocesses. In addition, $\exp (-\Omega)$ is the probability that no inelastic soft interaction in the re-scattering eikonal chain results in inelasticity of the final state at energy $s$ and inpact parameter $b$. 

As the $b$-dependences of single and double-diffractive dissociation amplitudes are distinct, the survival factor is different for each case. The survival factor decreases with energy due to the growth of the opacity, whereas increases with the slope $B$. In order to illustrate the calculation in Eq. (\ref{sdef}), let us consider the eikonal model (one channel) and assume a Gaussian $b$-dependence for the scattering amplitude for a hard process, ${\cal A}_H(s,b)\propto \exp (-b^2/R_H^2)$, with a constant hard radius $R_H$. The opacity can be oversimplified in the form $\Omega (s,b)=\frac{\sigma_0s^{\Delta}}{\pi R^2(s)}\exp [-\frac{b^2}{R^2(s)}]$. Here, $\Delta = \alpha_{\pom}(0)-1$ and the soft slope is $R^2(s)\propto \ln (s)$. In this simple case, Eq. (\ref{sdef}) can be analytically evaluated and then give the following: $<\!\!|S|^2\!\!>=\bar{R}_H \,\gamma [\bar{R}_H,a(s)]/[a(s)]^{\bar{R}_H}$. We use the notation $\bar{R}_H(s) \equiv \frac{R^2(s)}{R_H^2}$, $a(s)=\sigma_0s^\Delta$ and $\gamma (a,x)$ is the incomplete Euler gamma function \cite{GLMrev}. This rough calculation shows the model dependence on the input values $R_H$, $R(s)$ and $a(s)$. We quote Ref. \cite{GLMrev} for extensive review on the determination of LRG survival probability gap factor using the multi-channel eikonal models.

Concerning the model dependence, the single channel eikonal model considers only elastic rescatterings, whereas the multi channel one takes into account also inelastic diffractive intermediate re-scatterings. In general, current works in literature consider a more elaborate two or three channel eikonal model. These approaches describe correctly the relevant observables in $pp$ and $p\bar{p}$ collisions as $\sigma_{tot,\,el}$, $B_{el}$, $\sigma_{sd}$ and so on \cite{GLM,KMRsoft}. The corresponding survival probabilities of single, double and central channels are not identical because each one has a different hard radius. The available experimental observables which can be compared to the theoretical predictions of the survival probability factor are the hard LRG di-jets data obtained in the Tevatron and HERA \cite{GLMrev,KKMR}  as well as diffractive hadroproduction of heavy bosons ($W^{\pm}$ and $Z^0$) in the Tevatron \cite{GDMM}. A direct information on the survival probability factor is obtained from the diffractive hard jets ratio mesured in Tevatron, where $\frac{<\!|S|^2\!>(\sqrt{s}=0.63 \,\mathrm{TeV})}{<\!|S|^2\!>(\sqrt{s}=1.8\, \mathrm{TeV})}=2.2\pm 0.8$. The current theoretical predictions are in agreement with this measurement.

For our purpose, we consider the theoretical estimates  from Ref. \cite{KKMR} (labeled KMR), which considers a two-channel eikonal model that embodies pion-loop insertions in the pomeron trajectory, diffractive dissociation and rescattering effects. The survival probability is computed for single, central and double diffractive processes at several energies, assuming that the spatial distribution in impact parameter space is driven by the slope $B$ of the pomeron-proton vertex. We will consider the results for single diffractive processes with $2B=5.5$ GeV$^{-2}$ (slope of the electromagnetic proton form factor) and without $N^*$ excitation, which is relevant to a forward proton spectrometer (FPS) measurement. Thus, we have $<\!|S|^2\!>_{\mathrm{KMR}}^{\mathrm{SD}}=0.15,\,[0.09]$ and  $<\!|S|^2\!>_{\mathrm{KMR}}^{\mathrm{DPE}}=0.08,\,[0.04]$ for $\sqrt{s}=1.8$ TeV (Tevatron) [$\sqrt{s}=14$ TeV (LHC)]. There are similar theoretical estimates, as the GLM approach \cite{GLMrev}, which also consider a multiple-channel eikonal approach. We verify that those results are consistent with each other.  We quote Ref. \cite{GLMrev} for a detailed comparison between the two approaches and further discussion on model dependence.

For sake of illustration, we have adjusted the gap survival probability factor (for single diffraction) as a function of center-of-mass energy using the following functional form  $<\!|S|^2\!>^{\mathrm{SD}}(\sqrt{s})=a/(b+c\,ln^d\sqrt{s})$, in the interval $10\,\mathrm{GeV}\leq \sqrt{s}\leq 14\,\mathrm{TeV}$. This functional form is inspired in the fact that upon reaching the unitarity limit the fraction of diffractive events is expected to vanish as $1/\ln \sqrt{s}$ \cite{gspb}. For energies lower than $\sqrt{s}\leq 540$ GeV the estimates from Ref. \cite{BH} are used. The following parameters are obtained: $a=2.062$, $b=4.937$, $c=4.2\cdot 10^{-3}$ and $d=3.793$. This is useful to obtain the single diffractive cross sections as a function of energy when they are corrected by multiple-Pomeron scattering.

%
%

\section{Results and Discussion}
\begin{figure}[t]
\includegraphics[scale=0.47]{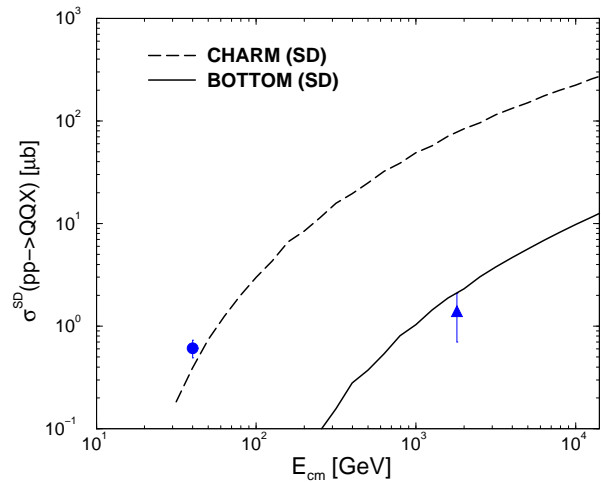}
\caption{(Color online) Single diffractive cross section for charm and bottom hadroproduction considering multiple-Pomeron corrections as a function of energy. A comparison to experimental data is also shown (see text).}
\label{fig:3}
\end{figure}

\begin{table}[t]
\caption{\label{tab:table1} Model predictions for single (SD) and central (DPE) diffractive  heavy flavor production in Tevatron and LHC. Numbers between parenthesis represent the estimates using the single Pomeron exchange.}
\begin{ruledtabular}
\begin{tabular}{lcll}
$\sqrt{s}$  & Heavy Flavor   & $R_{\mathrm{SD}}$ (\%) & $R_{\mathrm{DPE}}$ (\%)\\
\hline
1.96 TeV &  $c\bar{c}$  & $3.5 \,(23.4)$   & $0.08\, (1.2)$\\
1.96 TeV &  $b\bar{b}$  & $2.5 \,(16.9)$  & $0.05\,(0.6)$\\
1.96 TeV &  $t\bar{t}$  & $1.5\cdot 10^{-3}\,(0.03)$   & -----\\
14 TeV &    $c\bar{c}$  & $2.8\, (30.7)$   & $0.08\, (2.0)$\\
14 TeV &    $b\bar{b}$  & $2.1\,(23.5)$  & $0.035\, (0.9)$\\
14 TeV &    $t\bar{t}$  & $0.4\,(4.8) $   & $8\cdot 10^{-3}\, (0.23)$\\
\end{tabular}
\end{ruledtabular}
\end{table}

In what follows, we present  predictions for hard diffractive production of
heavy flavors based on the previous discussion.  In the numerical calculations, we have used the new H1 parameterizations for the diffractive pdf's \cite{H1diff}. The `H1 2006 DPDF Fit A' is considered, whereas a replacement by `H1 2006 DPDF Fit B' keeps the results unchanged. For the  pdf's in the proton we have considered the updated MRST parameterization \cite{mrst2004nlo}. In addition, we have used the cut $x_{\pom}<0.1$. The single Pomeron results, given by Eq. (\ref{sdxsect}), are presented in Figs. \ref{fig:1} and \ref{fig:2} for charm and bottom production (dot-dashed curves). The single diffractive contribution is large, being of order 15--20 \% from the total cross section. For the top case, which is not shown in a plot, we get a small cross section. Namely, $\sigma_{t\bar{t}}^{\mathrm{SD}}= 1.5\cdot 10^{-3}$ pb at $\sqrt{s}=1.96$ TeV and $\sigma_{t\bar{t}}^{\mathrm{SD}}= 100$ pb for $\sqrt{s}=14$ TeV. The cross sections for each heavy flavor rise with energy and differ approximately by a factor proportional to $1/m_Q^2$.  The results corrected by unitarity suppresion are a factor about $1/10$ lower than the single-Pomeron ones. In Fig. \ref{fig:3} we present the numerical results for the single diffractive production of charm (long-dashed curve) and bottom (solid curve) considering the suppression factor. We have multiplied Eq. (\ref{sdxsect}) by the energy dependent factor $<\!|S|^2\!>^{\mathrm{SD}}(\sqrt{s})$, as discussed in previous section. Of course, the present curves have to be considered as average results, since the absolute cross sections are sensitive to different choices for the quark masses and renormalization scale.

The results using the multiple-Pomeron correction can be somewhat compared with experimental estimates for the single diffractive cross section. However, data for diffractive production of heavy flavors are scarce.  For charm, we quote the most accurate data for production of $D^{*-}$ from the E690 experiment at Fermilab at $\sqrt{s}=40$ GeV \cite{sdccdata}. The cross section is $\sigma(pp\rightarrow p[c\bar{c}]\,X)=0.61\pm0.12(stat)\pm0.11(syst)$ $\mu$b, which is integrated over $x_F>0.85$. This data point  is represented by the filled circle in Fig. \ref{fig:3}. For the bottom case, the estimation $\sigma(pp\rightarrow p[b\bar{b}]\,X)=1.4\pm 0.7$ $\mu$b at $\sqrt{s}=1.8$ TeV is taken from Ref. \cite{Kop}. This value is obtained using theoretically predicted inclusive cross section $140$ $\mu$b and the experimentally measured  fraction $R_{b\bar{b}}$ of diffractively produced beauty \cite{sdbbdata}. This value  is represented by the filled triangle in Fig. \ref{fig:3}. Our curves reasonably describe these scarce data on diffractive production of heavy flavored hadrons.

Concerning the DPE cross sections, we verify that the  single-Pomeron predictions are still large. For instance, using Eq. (\ref{ddxsect}) one has  $\sigma_{c\bar{c}}^{\mathrm{DPE}}= 58$ $\mu$b and  $\sigma_{b\bar{b}}^{\mathrm{DPE}}= 1.2$ $\mu$b at $\sqrt{s}=1.96$ TeV. The top production is suppressed due to kinematic threshold related to the cut-off in the Pomeron spectrum, $x_{\pom}\leq 0.1$. This gives little room to observe diffractive top quark events at the Tevatron but is promising for the LHC, where the heavy flavour threshold suppression is less severe. The values reach $\sigma_{c\bar{c}}^{\mathrm{DPE}}= 460$ $\mu$b, $\sigma_{b\bar{b}}^{\mathrm{DPE}}= 13$ $\mu$b and $\sigma_{b\bar{b}}^{\mathrm{DPE}}= 5$ pb for $\sqrt{s}=14$ TeV. The results corrected by multiple-Pomeron suppression factor are a factor about $1/100$ lower than the single-Pomeron ones.

Let us now compute the diffractive ratios. The single diffractive ratio is defined as $R_{\mathrm{SD}}= \sigma^{\mathrm{SD}}/\sigma_{tot}$ and the central diffractive ratio by $R_{\mathrm{DPE}}= \sigma^{\mathrm{DPE}}/\sigma_{tot}$.   The results  are summarized in Table I, where the diffractive ratios for heavy flavor production  are presented for Tevatron and LHC energies. The multiple-Pomeron correction factor is taken from KMR model. The numbers between parenthesis represent the single Pomeron calculation. Based on these results we verify that the charm and beauty production in  single diffractive process could be observable in Tevatron and LHC, with a diffractive ratio of order 2--3 \%. The top production case is problematic, mostly due to the kinematic threshold. The predictions for central diffractive scattering are still not very promising. However, the study of these events is worthwhile, at least for charm and bottom production. The predictions can be taken as a guide since a more detailed investigation with detector acceptance is deserved. For illustration, in Fig. \ref{fig:4} the energy dependence of the single diffractive ration for charm and bottom is presented in the interval $0.63\leq \sqrt{s}\leq 14$ TeV. The ratios decrease on energy due to the gap survival probability factor which behaves as $1/\ln \sqrt{s}$.

\begin{figure}[t]
\includegraphics[scale=0.47]{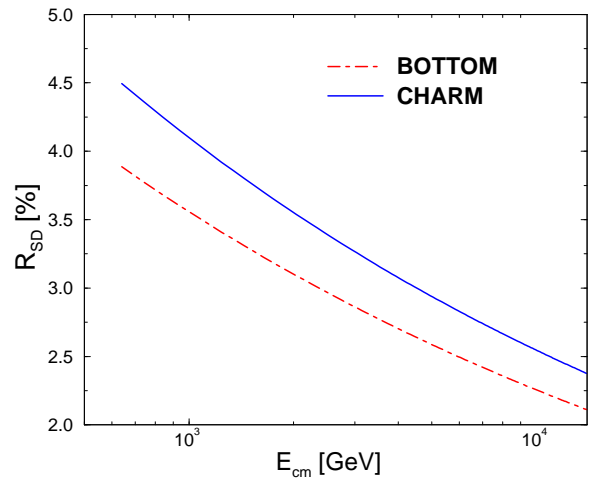}
\caption{(Color online) The single diffractive ratio, $R_{\mathrm{SD}}(\sqrt{s},m_Q)$, for charm and bottom hadroproduction considering multiple-Pomeron corrections as a function of energy.}
\label{fig:4}
\end{figure}

Our calculation can be compared to available literature in diffractive heavy flavor production. For instance, in a previous calculation presented in Ref. \cite{Heyssler}, the author has considered the leading order approximation for the hard scattering and old parameterizations for the Pomeron structure function. Corrections for multiple-Pomeron exchange were not considered. Ours results for the absolute cross sections are not comparable but the values for the diffractive ratios for the single Pomeron case are similar (at least for charm and bottom). The present calculation is in agreement to the recent investigations within the light-cone dipole approach \cite{Kop}. Our results for the single diffractive production of charm and bottom are similar to those in Ref. \cite{Kop}, including the overall normalization and energy behavior. For the top case, our results seem to be smaller than the dipole calculations.

In summary, we have presented predictions for diffractive heavy flavor production at the Tevatron and the LHC. In calculations rely on the Regge factorization (single-Pomeron exchange) supplemented by gap survival probability factor (correction for multiple-Pomeron exchange). For the Pomeron structure function, we take the recent  H1 diffractive parton density functions extracted from their measurement of $F_2^{D(3)}$. The results are directly dependent on the quark singlet and gluon content of the Pomeron. We did not observe large discrepancy in using the different fit procedure for diffactive pdf's. We estimate the multiple interaction corrections taking the theoretical prediction a  multiple-channel model (KMR), where the gap factor decreases on energy. That is,  $<\!|S|^2\!>\simeq 15$ \% for Tevatron energies going down to $<\!|S|^2\!>\simeq 9$ \% at LHC energy. The consideration of other models for the suppression factor does not introduce dramatic changes. We found that at the Tevatron single and central (DPE) diffractive charm and bottom quark production is observable with a single diffractive ratio $R_{q\bar{q}}^{\mathrm{SD}}$ between  3.5 \% (charm) and 2.5 \% (bottom). The DPE cross section for charm and bottom production lies above the total inclusive cross section for the top quark and might also be observable. The diffractive cross sections will be larger by at least one order of magnitude at the LHC and also diffractive top quark production could be observable. Therefore, the LHC can be a laboratory for diffractive scattering studies.



\section*{Acknowledgments}

This work was supported by CNPq, Brazil. The author thanks the hospitality of the Departamento de F\'{\i}sica Te\'orica of UFRJ (Brazil), where part of this work was performed. 


%
%


\end{document}